\documentclass[12pt,preprint]{aastex}

\def\ngc{NGC$\,$5252\ }

\newcommand{\Lsun}{L$_\odot$}
\newcommand{\Msun}{M$_\odot$}
\newcommand {\ia}{\'\i }

\slugcomment{v5 - 6 Jun 2002}

\shorttitle{The SED of NGC 5252}

\begin{document} 

\title{The IR spectral energy distribution of the Seyfert 2 prototype
NGC 5252}
\author{M. Almudena Prieto}
\affil{European Southern Observatory, D-85748 Garching, Germany}
\affil{Instituto de Astrofisica de Canarias, La Laguna, Spain}
\email{aprieto@eso.org}
\author{J.A. Acosta--Pulido}
\affil{Instituto de Astrofisica de Canarias, La Laguna, Spain}
\email{jap@ll.iac.es}

\begin{abstract}

The complete mid- to far- infrared continuum energy distribution
collected with the Infrared Space Observatory of the Seyfert 2
prototype NGC 5252 is presented.  ISOCAM images taken in the
3--15~{\micron\ } show a resolved central source that is consistent at
all bands with a region of
%(FWHM $\simeq$ 6.4") at the resolution of the observations (FWHM$\simeq$ 5") 
%and thus covering an 
about 1.3 kpc in size.  Due to the lack of on going star formation in
the disk of the galaxy, this resolved emission is associated with either dust
heated in the nuclear active region or with
bremsstrahlung emission from the nuclear and extended ionised gas. The
size of the mid-IR emission contrasts with the standard unification
scenario envisaging a compact dusty structure surrounding and hiding
the active nucleus and the broad line region.

The mid IR data are complemented with ISOPHOT aperture photometry in
the 25--200 \micron\ range.  The overall IR spectral energy distribution
is dominated by a well-defined component peaking  at $\sim 100$~\micron,
a characteristic temperature of
T$\simeq 20$~K and an associated dust mass of $2.5 \times
10^{7}$~\Msun which greatly dominates the total dust mass content of
the galaxy.  The heating mechanism of this dust is probably the
interstellar radiation field.  After subtracting the contribution of
this  cold dust component, the bulk of the residual emission is  attributed
to dust heated within the nuclear environment.  Its luminosity
consistently accounts for the reprocessing of the X-ray to UV emission
derived for the nucleus of this galaxy.  The comparison of \ngc spectral
energy distribution with current torus models favors large nuclear disk 
structure on the kiloparsec scale.

\end{abstract}


\section{Introduction}

\ngc is one of the best examples of  
anisotropy of the  nuclear radiation field. It exhibits 
a  perfect biconical morphology
of extranuclear ionized gas\citep{Tadhunter89}, extending out to
$\sim$~45"--50", equivalent to $\sim$~20~kpc\footnote{\ngc has  a redshift
${\rm z} \simeq 0.0230$, which gives a scale of $\simeq 433$~pc/arcsec, 
adopting H$_{0}=75~$km~s$^{-1}$~Mpc$^{-1}$.} and  
neutral HI gas  filling the regions outside the bicone of ionized gas
\citep{Prieto93}. Such selective distribution of gas 
implies an  intrinsically collimated radiation field 
at least at the extended narrow line region scale.

\ngc has been subjected to a wide range of multi-wavelength
observations from the X-ray to the radio domain. A large fraction of
the observations were aimed at revealing the nature and geometry of
the postulated obscuring material responsible for blocking the nuclear
light in directions outside  the bicone of ionized gas.  The
observational evidence points to a heavily obscured nucleus and the 
broad line region (BLR): a
band of red material lying across the nucleus and perpendicular to the
bicone of ionized gas  is seen in near IR -- optical
ratio maps of the galaxy \citep{Kotilainen95}.  
The presence of an obscured nucleus is also
confirmed in the high energy domain. The 0.4 -- 10 keV ASCA spectrum
of the galaxy \citep{Cappi96} reveals a heavily obscured nucleus with
N(H) $\rm \sim4.3\times10^{22}~cm^{-2}$.  This implies about 20 mag of
extinction in the visible, sufficient to absorb all the nuclear
optical and UV light in the line of sight.
On the other hand, broad permitted lines have been observed in 
the optical \citep{Acosta96, Osterbrock93} 
and  NIR range \citep{Ruiz94, Goodrich94}. 
The modeling of the nuclear spectral energy distribution (SED) and
line spectrum in  NGC 5252 is  consistent with the nuclear 
ionizing radiation being heavily absorbed  \citep{Contini98}. 

Because of the heavy central obscuration advocated by the above
observations, the mid to far IR region becomes an ideal window for
studying the nuclear emission in detail. The 20 mag extinction
measured in the optical reduces to about 1.5 mag at 10~\micron.  This
paper presents new photometric data covering the mid to far IR range
(from 3~\micron to 200~\micron).  Together with near-IR data available
from the literature, one of the most complete IR SED among Seyfert
galaxies is presented.  In contrast to most Seyfert galaxies, NGC 5252
shows no evidence for star--formation activity across its disk:
H$\alpha$ imaging reveals the distribution of ionized gas to be
restricted to the nucleus and the bicone only (Prieto \& Freudling
1996, Tsvetanov et al. 1996). This unique advantage makes of \ngc an
ideal target for pursuing ``clean'' studies of its central nuclear
emission.

\section{The data} 

Mid-IR images of NGC~5252 were obtained in several filters using the
ISOCAM instrument \citep{Cesarsky96} on board ISO\footnote{Based on
observations with ISO, an ESA project with instruments funded by ESA
Member States (especially the PI countries: France, Germany, the
Netherlands and the United Kingdom and with the participation of ISAS
and NASA.}.  A total of eight filters were used (see Table
\ref{ta:fluxes}): five centered on the continuum at 15, 14.5, 11.37,
6.75, and 3.72~{\micron\ } respectively and three in the PAH emission
at 3.3~\micron, CO line at 4.7~\micron\, and the broad Si absorption
feature at 9.63~\micron respectively.  
The  pixel field of view was 3
arcsec.  All images were collected following a micro--scanning pattern
of 3\,x\,3 points, and a step size of 2 arcsec in both x and y
directions. In this way the PSF was oversampled, which allows a
better detection of possible extended emission.  The data were
processed using the version 4.0 of CIA \citep{Ott97}. This was used for
dark signal subtraction, deglitching, drift correction (for the LW
camera) using \citet{Fouks95} formula, flat--fielding, flux
calibration, and mosaicing.  Unfortunately, the images in the 3.3 and
4.7 {\micron\ } filters (taken with the SW camera) are affected by
hysteresis that  could not be corrected.

Integrated aperture photometry in the
25 to 200~{\micron\ } range measured by ISOPHOT \citep{Lemke96} are
also presented.  The ISOPHOT data were reduced using PIA v9.1
\citep{Gabriel97} following the standard signal corrections
\citep{Laureijs00}\footnote{ The ISO Handbook, vol. V is available at:
http://www.iso.vilspa.esa.es/manuals/HANDBOOK/V/pht\_hb} except for
signal non-linearity, which was not applied.  The observation in
chopped mode at 25~{\micron\ } was calibrated by using a mean value of
detector responsivity corrected for systematic variation with orbital
position \citep{Abraham01}.  In the other cases, the internal
calibrators were used to determine the responsivity.  Due to the
faintness of the source, the fluxes at 60 and 90~{\micron\ } were
derived using only the central pixel of the ISOPHOT--C100 array, after
correction by signal losses due to partial covering of the PSF. These
values were re-calibrated by the scaling factors 0.8 and 1.27, at 60
and 90~\micron, respectively. These scaling factors were derived after
comparison of the fluxes as measured by ISOPHOT at those bands and
measured by IRAS at 60 and 100~{\micron\ } of a sample of CfA Seyfert
galaxies (Gonz\'alez--Hern\'andez et al., in preparation).  The IRAS
data at 60 and 100~{\micron\ } \citep{Edelson87} were also included 
in the analysis.
The fluxes measured at filters beyond 120~{\micron\ } were
re-calibrated after comparison of the background measurements with
respect to the corresponding values in the COBE/DIRBE maps.

\begin{deluxetable}{rrrll}
\tablewidth{0pt}
\tablecaption{Nuclear photometry of NGC 5252\label{ta:fluxes}}
\tablehead{\colhead{Filter}  & 
\colhead{$\lambda$} & \colhead{Ap.} &  \colhead{$F_\nu$} & \colhead{Ref.} \\
\colhead{} & \colhead{[\micron]} & \colhead{[arcsec]} & \colhead{[mJy]} & \colhead{}}
\startdata 
  5230\AA  &  0.52 &  3   & 1.1    & 1 \\
  5230\AA  &  0.52 &  nuc & 0.012  & 1 \\
  7027\AA  &  0.70 &  3   & 2.7    & 1 \\
  7027\AA  &  0.70 &  nuc & 0.053  & 1 \\
  J        &  1.25 &  3   & 11.0   & 1 \\
  J        &  1.25 &  nuc & 11.0   & 1 \\
  H        &  1.65 &  3   & 15.0   & 1 \\
  H        &  1.65 &  nuc & 2.6    & 1 \\
  K        &  2.2  &  3   & 13.0   & 1 \\
  K        &  2.2  &  nuc & 3.78   & 1 \\
  CAM-SW6  & 3.72  &  15  &  16.6 & 2 \\
  CAM-LW5  & 6.75  &  15  & $20\pm5$ & 2 \\
  CAM-LW7  & 9.63  &  15  & $34\pm6$ & 2 \\
  CAM-LW8  & 11.37 &  15  & $48\pm6$ & 2 \\
  IRAS-12  & 12    &  15  &  40:  & 3 \\
  CAM-LW3  & 14.5  &  15  &  $47\pm7$ & 2  \\
  CAM-LW9  & 15    &  15  &  $63\pm8$ & 2  \\
  PHT-P25  & 25   &   60  & $75.0\pm60$ & 2  \\
  IRAS-25  & 25   & 	  & 56:  & 3 \\
  IRAS-60  & 60   &  90   & $425\pm55$ & 3  \\
  PHT-C60  & 60   &  60   & $380\pm110$ & 2  \\
  PHT-C90  & 90   &  70   & $498\pm150$ & 2   \\
  IRAS-100 & 100  &  90   & $750\pm130$  & 3  \\
  PHT-C120 & 120  & 	  & $840\pm70$ & 2  \\
  PHT-C135 & 150  & 	  & $1060\pm50$ & 2  \\
  PHT-C180 & 180  & 	  & $710\pm50$ & 2 \\
  PHT-C200 & 200  & 	  & $570\pm160$ & 2 \\
  CO(2-1)  & 1332  &	  & 27: & 4 \\
  CO(1-0)  & 2677  &	  & 21: & 4 \\
\enddata
\tablecomments{Colons indicate an upper limit.}
\tablerefs{1 - \citet{Kotilainen95}; 2 - this article; 3 - \citet{Edelson87};
 4 - \citet{Prieto96}.}
\end{deluxetable}

\section{Results}

\subsection{Extended  Mid IR Emission}

The IR emission of \ngc in the ISOCAM images (3-15um range) shows
concentrated to the nuclear region; yet, the emission is however
marginally extended in all the filters. The extended nature of the 
emission was assesed in the two following ways. First, 
each image was compared with the modeled ISOCAM point spread function
(PSF) corresponding to each filter \citep{Okumura98}\footnote{ISOCAM
PSF report is available at
http:www.iso.vilspa.esa.es/users/expl\_lib/CAM\_list.html}.  In all 
cases, the  galaxy profile was found broader than the theoretical PSF.  
Second, a comparison with a measurement of a calibration star 
observed with exactly the same raster configuration as that used for
\ngc was used.  This comparison could only be done at 15\micron\
(LW3 filter) as no other stars with the same observation configuration
exist in the ISO archive (Fig. 1).  This comparison was conclusive. 
The measured FWHM of the galaxy profile at
15$\micron$ is 5.6" whereas that of the stellar profile is 4.6"
(Fig. 1).  After de-convolving for the PSF width, {\it an extension of
$\simeq 3.3$~arcsec consistent in all filters is found.}\footnote{There are slight discrepancies between the theoretical PSF and that of the star profile which affect the peak of the function and the wings (see
discussion by Okumura, 1998).  However the difference between the FWHM
of both PSFs never exceeds 10\% for the filters used here.}
The compactness of the emission in the ISOCAM images
hampers any assessment on  the morphology definition. 

Previous claims regarding the presence of obscuring structure in the
circumnuclear region of this galaxy are reported by
\citet{Kotilainen95}: a redder nuclear band, perpendicular to the
ionized gas bicone and extending $\sim 5"$ E--W across the nucleus was
found in near-IR optical colour maps.  \citet{Tsvetanov96}, on the
basis of HST optical observations reported on a D-shaped obscuring
structure N-E of the nucleus with a size of $\sim 3"$ was reported.
Thus, although the morfology of the mid-IR emission is undetermined,
its size is consistent with that of the reported obscuring nuclear
structures seen in the near-IR and optical images.
 
There are few Seyfert galaxies in which the mid-IR emission is known
to be extended.  In NGC~1068, the emission from 7.9 to 25~{\micron\ }
extends over few arcseconds, aligning with the radio jet and roughly
correlating with the [OIII] morphology
\citep{Cameron93,Bock00}. \citet{Bock00} found that about 2/3 of the
total emission is contained in an unresolved central core (0.1"
resolution) and the other 1/3 in the extended emission.
\citet{Krabbe01} report on the detection of extended mid IR emission
in Circinus and NGC~3281.

\begin{figure}
\epsscale{0.9} \plotone{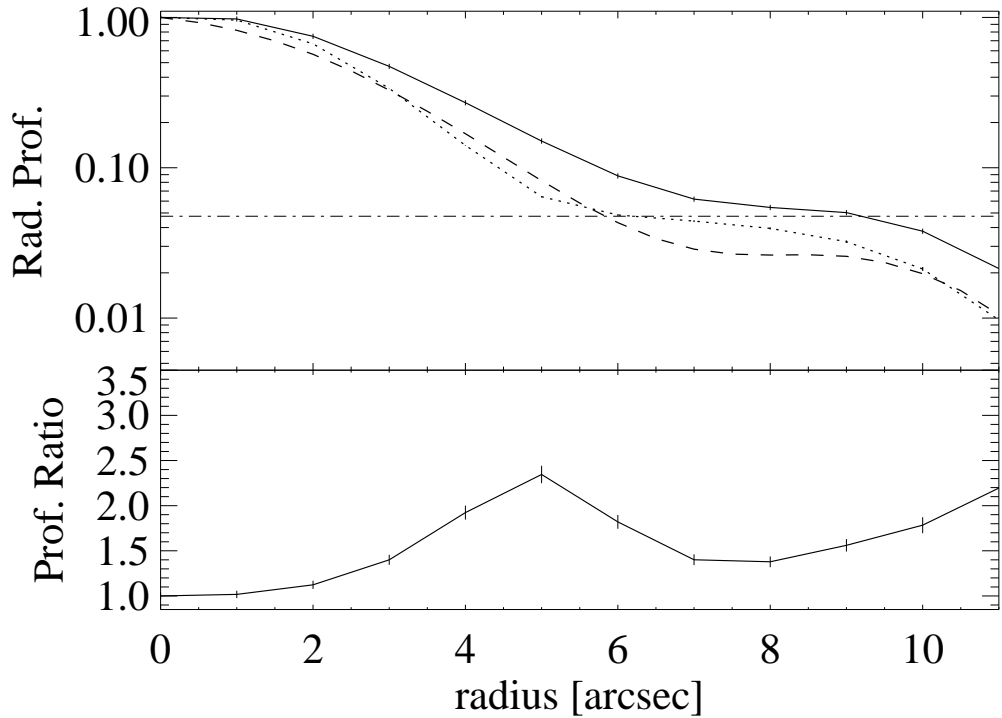}
\caption{Normalized radial intensity profiles of NGC 5252 at
15~\micron\ (continuous line). For comparison, the theoretical (dashed
line) and the observed (dotted line) PSFs are also shown. The
horizontal line represents the detection limit as measured by the
$3\sigma$ of the image background.  The ratio between the NGC 5252
radial profile and the observed PSF is represented in the bottom
panel. The difference between the theoretical and the observed PSFs at
the inner radii is due to the pixel sampling. }
\label{fi:prof}
\end{figure}

\subsection{IR Spectral Energy Distribution}
\label{se:IRSED}

\begin{figure}
\epsscale{0.9} \plotone{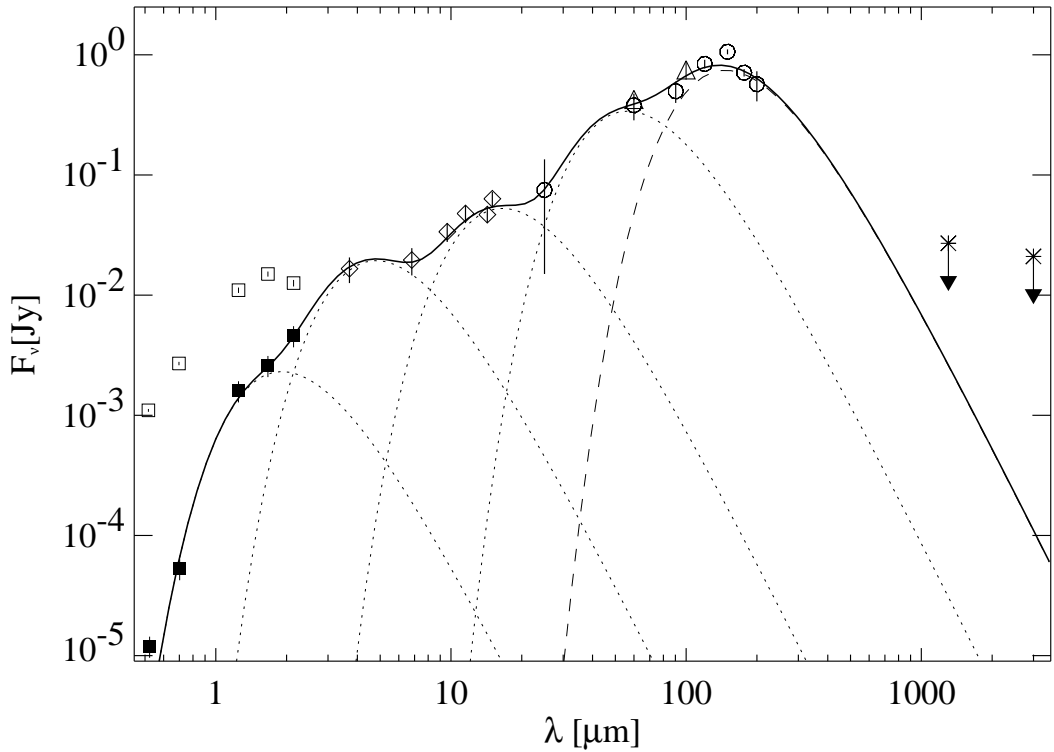}
\caption{IR Spectral energy distribution of NGC 5252.  Circles
represent ISOPHOT photometric data; diamonds,  ISOCAM data; 
triangles, IRAS data. For sake of clarity,  IRAS upper limits at 
12 and 25\micron\ are not inlcuded.
The open and filled squares represent the 3" aperture and
stellar-subtracted nuclear fluxes, respectively from
\citet{Kotilainen95}.  Two upper limits in the
submillimetric range are included \citep{Prieto96}.  The continuous
line is a parametrization of the SED as the sum of five modified
blackbodies with T\,=\,20, 50, 175, 600, and 1500~K.  The coolest
component is represented by the dashed line and the other blackbody
components as dotted lines. The SED shows its maximum at
$\sim 100$~\micron.}
\label{fi:sed}
\end{figure}

The IR spectral energy distribution of \ngc including the new ISOCAM
and ISOPHOT data is shown in Fig.~\ref{fi:sed}.  The
stellar-subtracted nuclear continuum fluxes in the near IR to optical
range by \citet{Kotilainen95} are also shown.  A clear feature in the
SED is the prominent bump enclosing the 60 to 200~\micron\ range and
dominating the continuum emission.

Up to IRAS, the bulk of IR emission in active galaxies has in general
being associated with dust reprocessing of the intense UV--optical
nuclear radiation. Although still lacking spatial resolution, the
wider wavelength coverage and better sampling of the IR continuum by
ISO has allowed \citet{Perez01} to discern three main IR components in
the SED of Seyfert galaxies. These are a very cold component (T~$\sim
15 - 20$~K), corresponding to dust heated by the interstellar
radiation field; a cold component (T~$\sim 40 - 50$~K), which is
associated with star forming regions; and a warm component (T~$\sim
100 - 200$~K) associated with dust heated by the active nucleus and/or
nuclear starbursts.  As our IR wavelength coverage for NGC 5252 gets
up to the near-IR, a simple parametrization of its SED as the sum of a
minimum number of modified blackbodies was introduced.  This
parametrization has  the purpose of getting  a characteristic value of the
luminosity, temperature and mass of the typical dust.
Fig.~\ref{fi:sed} shows the fit to the SED as the sum of a minimum
five modified blackbodies with T\,=\,20, 50, 175, 600, and
1500~K\footnote{We have also tried the inversion method described by
\citet{Perez98}. Their method yields seven components with the
following temperatures T\,=\,20, 50, 133, 182, 380, 820 and 1500~K.
The differences between our decomposition and theirs appear at the
gaps in the SED (2 to 7~\micron\ and 25 to 60~\micron), however the
main components at 20 and 50~K are preserved.}.  It should be stressed
that regardless of this parametrization, the local bump at
90--200~\micron\ is by itself well defined by by a single modified
blackbody with T$\sim 20 K$. We will single out this component as a
different one from the rest of the IR SED. The arguments are as
follows.

\citet{Danese92} shown that most of the far IR (IRAS at 60 and
100~\micron) emission in Seyfert galaxies is unrelated to nuclear
activity.  \citet{Perez01} and \citet{Acosta01} reached the same
conclusion on the basis of a much wider and better sampled spectral
energy range covered by the 16--200~\micron\ ISOPHOT data.  Recently
\citet{Prieto01, Prieto02} show that the X-ray emission and the
coronal line emission in Seyfert galaxies, both being typical
indicators of the nuclear activity, are correlated with the mid-IR
emission but unrelated to the far IR emission.  In the case of NGC
5252, the lack of star formation in the disk of the galaxy (sect. 1),
the absence of dust lanes or knots across its disk, as evidenced in
optical-IR colour maps \citep{Kotilainen95}, the compactness of the
emission in the 3--15\micron\ range, lead us to argue that the bulk of
the near- to mid- IR emission is of nuclear origin whereas the far-IR
emission, represented by the 20 K component, is a separate component
unrelated to the nuclear activity. This component is  
associated with the galaxy as a
whole and is most probably due to dust heated uniformly across the galaxy by
the interstellar radiation field.

The luminosity, temperature  and dust mass associated with the 
galaxy component (20 K) 
and with the nuclear component (sum of the other
blackbodies) was estimated following \citet{Kennicutt98} and is
provided in Table 2.

\begin{deluxetable}{lcccc}
\tablewidth{0pt} \tablehead{ \colhead{Comp.}  & \colhead{Temp} &
\colhead{$F_{{\rm IR}}$} & \colhead{$L_{{\rm IR}}$} &
\colhead{M$_{dust}$} \\ & \colhead{[K]} &
\colhead{[erg~cm$^{-2}$~s$^{-1}$]} & \colhead{[erg~s$^{-1}$]} &
\colhead{[\Msun]}} \tablecaption{Derived IR properties of
\ngc\label{ta:lumymas}} \startdata Galaxy & 20$\pm$4 & $(1.7 \pm
0.5)\times 10^{-11} $ & $(1.8 \pm 0.5)\times 10^{43} $ & $(2.5 \pm 1.0
)\times 10^7$ \\
%Cold      & 51$\pm$14    & $(2.1 \pm 0.5)\times 10^{-11}$  & 
   % $(2.2 \pm 0.5)\times 10^{43} $ & $(1.0 \pm 0.4) \times 10^5$ \\
%Warm      & 180$\pm$40   & $(1.1 \pm 0.2)\times 10^{-11} $  & 
   % $(1.2 \pm 0.3)\times 10^{43} $ & $33 \pm 11 $ \\
%Hot       & 600$\pm$170  & $(1.5 \pm 0.5)\times 10^{-11} $  & 
   % $(1.5 \pm 0.5)\times 10^{43} $ & $\ll 1$ \\
%Very hot  & 1500$\pm$300 & $(3.9 \pm 1.0)\times 10^{-12} $  & 
   % $(4.1 \pm 1.1))\times 10^{42} $ & $\ll 1$ \\ 
Nuclear & 50 -- 1500
& ($5.0 \pm 0.8)\times 10^{-11}$ & $(5.3 \pm 0.8)\times 10^{43} $ &
$1.0 \times 10^5$ \\ 
Total & ...  & $(6.2 )\times 10^{-11}$ &
$(6.5 )\times 10^{43} $ & $(2.5 \pm 1.0 )\times 10^7$ \\ \enddata
\end{deluxetable}

\section{Discussion}

\subsection{Origin of the Extended Mid IR emission}

The size of the extended mid-IR emission detected in \ngc corresponds
to a emitting region of $\sim$1.3~kpc. 
%This is larger than the usually
%few pc scale region envisaged in torus models for Seyfert galaxies
%(\citet{Pier93}; \citet{Granato94}).

Accordingly to the previous SED decomposition, dust emitting at the
mid IR waves should reach temperatures $\simeq 175$~K (see also
\citet{Perez01}.  The luminosity associated with this warm component
is $1.1\times 10^{-11}$~erg~cm$^{-2}$~s$^{-1}$.
%%(see Table\ref{ta:lumymas}). 
Most of this emission should come from the nucleus but not necessarily
all. For example, in NGC~1068, 1/3 of the 10\micron\ emission is
extended across the nucleus.  Some plausible mechanisms for heating
the dust to such temperatures outside the nucleus include intense star
formation, shocks, and the nuclear radiation.

In NGC 5252, a circumnuclear starburst can be ruled out
(cf. Sect. 1). In addition, the colours of the galaxy in the
circumnuclear region are compatible with those of a passively evolved
elliptical with moderate extinction \citep{Kotilainen95}.

Heating by the nuclear radiation is not trivial if thermal equilibrium
between dust grains and the nuclear radiation has to prevail.  The
nuclear luminosity of NGC 5252 is estimated $Lf\sim (0.8 - 6.7)\times
10^{10}$~\Lsun \citep{Kotilainen95}, the range reflects the variation
in power-law spectral index, f is the covering factor. For $T=175$~K,
the dust should be placed to a maximum distance from the nucleus of
$R=12-30$~pc. This is a factor 50 to 20 below the size of the emission
at 15\micron, $\sim 650~pc$ radius.  The nuclear luminosity may be
greatly underestimated if the covering factor of the ionized gas is
much lower than unity.  Still, to increase R by a factor of 10 the
luminosity should increase by factor 100.  However, most important is
the fact that the presumed thermal equilibrium cannot be applied to
very small grains and PAHs which are transiently heated to high
temperatures by a single photon absorption, later re--emitting their
energy in the mid IR.  It is worth mentioning that the extended mid
-IR emission in NGC~1068 does not present a radial variation of the
color temperature as expected for a uniform distribution of dust
grains \citep{Bock00}. That could be explained in terms of very small
grains transiently heated by the nuclear radiation field.  The
presence of PAHs in NGC 5252 cannot be confirmed from the present
data.  PAH emission is however claimed to be a general feature in the
spectrum of Seyfert 2 galaxies \citep{Clavel00}; yet, this emission is
most probably linked to circumnuclear star--forming regions though.
Interestingly, to account for the near--IR emission in NGC~1068,
\citet{Efstathiou96} propose the existence of optically thin dust
distributed within the ionization cone.  The extended mid IR emission
in \ngc may have similar origin.

Dust could also be heated by shocks, associated e.g., with the
propagation of the radio jet.  The modeling of the nuclear SED and line spectrum requires of shock excitation coupled with  photoionization to
provide a  self-consistent account of the
X-rays to radio properties of this galaxy  \citet{Contini98}.
Shocks however do not appear to play a key role in the extended emission line
spectrum of the galaxy which Contini et al found it to be dominated by
the nuclear radiation only. Therefore, it seems improbable that the
extended mid-IR emission be due to shock-heated dust.

An alternative possibility for the origin of the extended mid-IR is
 bremsstrahlung emission due the mandatory cooling process of the
 photoionised and shock heated gas.  The bremsstrahlung component
 peaks in the optical-UV range but its slowing decreasing tail,
 particularly in the near to mid IR range, can largely dominate
 the nuclear SED at these waves (see Contini \& Viegas 2000 for a
 thorough discussion). Thus,  the extended
 near- to mid- IR emission in NGC 5252, also perhaps in NGC 1068, may be 
 association with bremsstrahlung from the nuclear and extended
 ionized gas in the bicone.

\subsection{The nuclear IR emission}
\label{se:nucSED}

On the arguments provided in sect. 3.2, the IR SED of \ngc was
 separated into two main components: a nuclear component and a cold
 component (T$\sim$20 K) associated with the galaxy as a
 whole. Because of the well definition of the 20 K galaxy component,
 the nuclear contribution could be isolated by subtracting that galaxy
 component from the total emission.  The inferred IR luminosity for
 th residual component is $5.3 \times 10^{43}$~erg~s$^{-1}$ (see Table
 \ref{ta:lumymas}).  \citet{Kotilainen95} provided an estimate for the
 hidden, nuclear optical/UV luminosity expected to be reprocessed by
 circumnuclear dust to be $L_{{\rm repr}} \sim 2.6 - 8.5 \times
 10^{43}$~erg~s$^{-1}$ (range varies depending on the assumed
 power-law spectral index of the ionizing radiation).  
Despite the uncertainties, both independently derived
 luminosities are in  fair agreement and thus consistent with
 reprocessing of the nuclear radiation by circumnuclear dust.

The estimated mass of dust from the residual component is $\sim10^5
\, M_\odot$ (Table 2).  An alternative estimate of the mass of dust
 in the nucleus  can be derived from  extinction measurements.
\citet{Kotilainen95} found that the stellar galactic (in a 6"--12"
ring) and nuclear (within a 3" region) light suffer an extinction of
$A_V \sim 0.5$~mag and $A_V \sim 1$~mag, respectively. Assuming a
canonical dust to gas mass ratio, those values imply dust masses of
$6.7 \times 10^5$~{\Msun\ } and $1.2 \times 10^5$~\Msun respectively,
in good agreement with the value  for the nuclear component derived here 
(Table 2).

\begin{figure}
\epsscale{0.9} \plotone{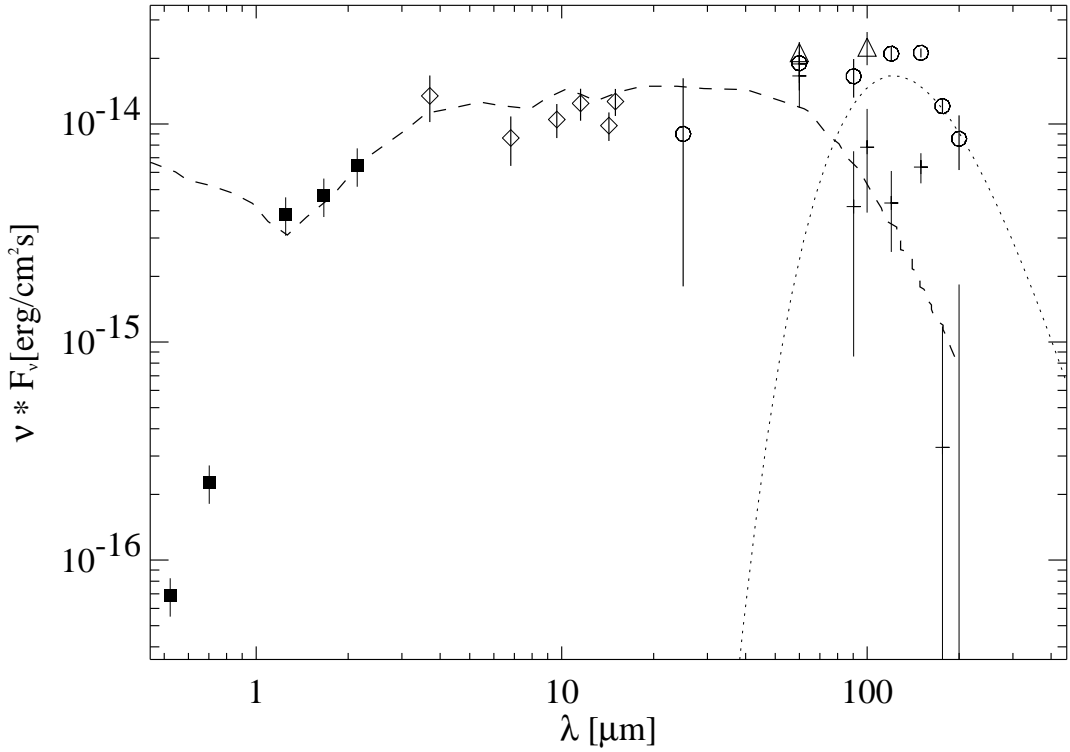}
\caption{Residual SED after removing the galaxy component.  Symbols
are as in Figure \ref{fi:sed}.
Simple crosses represent the far IR data after subtracting the 20~K
component.  The model proposed by \citet{Granato94} for NGC~1068 is
shown as dashed line.  Note the large width of the IR bump, about two
decades on wavelength.}
\label{fi:ene_nuc}
\end{figure}

The residual SED after removal of the galaxy component is shown in
Fig. \ref{fi:ene_nuc}.  First to notice is the width of the nuclear IR
bump, which covers about 2 decades in wavelength.  This
``nuclear'' SED is compared with current torus models next.  The best
compromise was found with \citet{Granato94} models.  These are flared
disk configurations characterized by a large radial extension, up to
hundreds of parsecs, and a moderate optical thickness.  Other models,
such as the compact torus proposed by \citep{Pier93} or the tapered
disks by Efstathiou et al. (1996), would need additional components to
account for the width of the bump and the near IR part of the SED.
Fig. \ref{fi:ene_nuc} shows the  proposed model for NGC~1068 by
\citet{Granato94} on top of \ngc nuclear SED.  The overall match is
remarkable, and indicates that models supporting large disk structures
should be further explored.
% \citet{Granato94} estimate a
%mass of $10^4 - 10^5 \, M_\odot$ for the flared torus, which is in
%fair agreement with the mass derived from the nuclear IR luminosity in
%NGC 5252 (see Table \ref{ta:lumymas}).

\subsection{Very Cold Dust Emission}

The IR luminosity and dust mass associated with the T~$\simeq 20$~K
component are given in Table \ref{ta:lumymas}.  This component
dominates the dust mass content of the galaxy. It is therefore worth
comparing it with alternative galaxy mass estimates and with those
derived from other galaxies of similar morphological type.

The dust mass derived from  extinction measurements in the
circumnuclear region of NGC 5252 (sect. 4.2) is about a factor 100
below that inferred from the galaxy component (Table 2).
Therefore, there is more dust in \ngc apart from that detected in the
circumnuclear region.  \citet{Prieto96} found HI emission distributed
outside the ionization cones and extending about 1 arcmin. A total
mass of neutral gas of $1.8 \times 10^9$~\Msun is derived ( an upper
limit for the molecular gas of $3.6 \times 10^8$~\Msun is derived from
millimetric observations).  If compared the HI mass with that  derived 
from the 20 K component, a gas to dust ratio of $\sim$ 98 is found, 
very close to
the canonical value.  This  suggests that the  cold dust
is probably related to the neutral gas and  most probably spread throughout the galaxy.

It is known since IRAS that early--type galaxies are not entirely
devoid of interstellar material \citep{Young89}, and \ngc is not an
exception.  \citet{Bregman98} studied the IRAS 60 and 100{\micron\ }
emission of a sample of normal elliptical and S0 and interpreted the
emission as due to  dust being extended throughout the galaxy and 
therefore, difficult to detect as obscuration patches.  Their derived dust
temperatures are likely overestimated since 100~{\micron\ } is not yet
on the falling turnover of the SED, hence the  dust masses
are expected to be underestimated. For comparative purposes,  the  temperature and
 mass of the dust in \ngc using the IRAS 60 and 100~{\micron\ } fluxes only, 
were derived. These yield
$T_{{\rm dust}} \sim 40$~K and M$\sim$ $1.4 \times 10^6$~\Msun respectively. 
Both values are well within the range  found in
Bregman's sample but close to the highest ones, which may be due to
the fact that \ngc harbors an active nucleus.

\section{Conclusions}

Based on new mid- and far- IR data collected with ISOCAM and ISOPHOT
instruments, a detailed IR SED of the Seyfert 2 prototype
\ngc is presented.  The main results are as follows.

\begin{itemize}
\item The IR SED of \ngc can be disentangled into two main components:
a nuclear component accounting for the bulk of the luminosity in the 1
-- 60~\micron\ range, and a  cold component (T$\simeq 20~K$),
 peaking at about100~\micron\ ,  associated with the galaxy as a whole.
The latter accounts for 25\% of the total IR luminosity
and almost the totality of the dust mass content of the galaxy.

\item The mid IR emission (6--15~\micron) detected in ISOCAM images is
concentrated toward the nuclear region of the galaxy.  This emission
is however resolved in all ISOCAM filters but its
morphology is undefined. A  limit for the size of the region is set
to 3.3 arcsec, $\sim$1.3 kpc diameter. This size
consistent with that  of the nuclear  obscuring structure  seeing in
optical and near-IR colour maps of the galaxy.  Possible origin for this
extended emission includes small dust grains 
transiently heated by the nuclear radiation 
or bremsstrahlung from nuclear and extended ionised gas.

\item Comparison of the SED with current torus models  favours
models supporting large scale disk structures. Indeed, the size of
\ngc central obscuring structure  and of the mid-IR region point out to
kpc-scale central disk.
\end{itemize}

\acknowledgements The COBE dataset were developed by the NASA Goddard
Space Flight Center under the guidance of the COBE Science Working
Group and were provided by the NSSDC. Ana P\'erez Garc{\ia}a provided
us with the results from the SED inversion method.

\end{document}